\documentclass[12pt]{article}
\usepackage{graphicx,amssymb,amsfonts}
\newlength{\extraspace}
\newlength{\extraspaces}
\newcommand{\be}{\begin{equation}
\addtolength{\abovedisplayskip}{\extraspaces}
\addtolength{\belowdisplayskip}{\extraspaces}
\addtolength{\abovedisplayshortskip}{\extraspace}
\addtolength{\belowdisplayshortskip}{\extraspace}}
\newcommand{\ee}{\end{equation}}

\newcommand{\ba}{\begin{eqnarray}
\addtolength{\abovedisplayskip}{\extraspaces}
\addtolength{\belowdisplayskip}{\extraspaces}
\addtolength{\abovedisplayshortskip}{\extraspace}
\addtolength{\belowdisplayshortskip}{\extraspace}}
\newcommand{\ea}{\end{eqnarray}}
\newcommand{\nonu}{\nonumber \\[.2mm]}
\newcommand{\A}{&\!\!\!}

\begin{document}

\begin{center}
{\bf  Spherically symmetric  solutions on non trivial frame in
$f(T)$ theories of gravity}

\end{center}
\begin{center}
{\bf Gamal G.L. Nashed}
\end{center}
\centerline{\it Mathematics Department, Faculty of Science, King
Faisal University, P.O. Box.} \centerline{\it 380 Al-Ahsaa 31982,
the Kingdom of Saudi Arabia\footnote{\small{ Mathematics Department,
Faculty of Science, Ain
Shams University, Cairo, 11566, Egypt \vspace*{0.1cm}\\
Center for Theoretical Physics, British University of  Egypt
 Sherouk City 11837, P.O. Box 43, Egypt
\vspace*{0.1cm} \\
 Egyptian Relativity Group (ERG) URL:
http://www.erg.eg.net}}}
\bigskip
\centerline{ e-mail:nashed@bue.edu.eg}

\hspace{2cm} \hspace{2cm}
\\
\\
\noindent{\narrower\small\sl{New solution  with  constant torsion  is derived using the field equations of $f(T)$.
Asymptotic form of  energy density, radial and transversal pressures are shown to
met  the standard energy conditions. Other solutions are obtained for
vanishing  radial pressure and
physics relevant to the resulting models are discussed.}
\par}\vskip 3mm\normalsize

\noindent{\narrower\sl{PACS:  04.50.Kd,  04.70.Bw, 04.20. Jb}
{\rm\hspace*{13mm}DOI:}
\par}\vskip 3mm


Recent observational data suggest that our universe is
accelerating.$^{ \rm {[1,2]}}$  This acceleration is explained in
terms of the so called dark energy (DE).
DE could also result from a cosmological constant, from an ideal
fluid with a different shape of equation of state and negative
pressure$^{ \rm {[3]}}$, etc.  It is not clear what
type of DE is more seemingly to explain the current era of the
universe.  A very attractive
possibility is the already mentioned as ``modification of General
Relativity" (GR). Amendments  to the Hilbert-Einstein action through the
introduction of different functions of the Ricci scalar have been
systematically explored by the so-called $f(R)$ gravity models,
which reconstruction has been developed.$^{ \rm {[4-9]}}$

In recent times, a new attractive modified gravity to account for the accelerating
expansion of the universe, i.e., $f(T)$ theory, is suggested  by
extending the action of teleparallel gravity$^{ \rm {[8,10-13]}}$  similar to the $f(R)$ theory, where T is the torsion
scalar.   It has been demonstrated that the
$f(T)$ theory can not only explain the present cosmic acceleration
with no need to dark energy$^{ \rm {[14]}}$ , but also provide an
alternative to inflation without an inflation$^{ \rm {[15,16]}}$ . Also it is shown that $f(T)$
 theories are not dynamically
equivalent to teleparallel action plus a scalar field under conformal transformation.$^{ \rm {[17]}}$  It
therefore has attracted some attention recently. In this regard,
Linder$^{ \rm {[18]}}$  proposed two new $f(T)$ models to explain the
present accelerating expansion and found that the $f(T)$ theory
can unify a number of interesting extensions of gravity beyond
general relativity.$^{ \rm {[19]}}$

The objective of this work is to find spherically symmetric
 solutions, under the framework of $f(T)$,  using anisotropic
spacetime. In \S 2, a brief review of $f(T)$ theory is presented.
In \S 3,  non trivial spherically symmetric spacetime is
provided and  application  to the field equation of $f(T)$ is
done. Several new spherically symmetric anisotropic solutions are derived in
\S 3. Several figures to demonstrate  the asymptotic behavior of  energy density and
transversal  pressure are  also given in \S 3. Final section is devoted
to the key results.

In a spacetime with absolute parallelism  parallel vector
fields ${h_a}^\mu$$^{ \rm {[20]}}$  identify the nonsymmetric
connection \be {\Gamma^\lambda}_{\mu \nu} \stackrel{\rm def.}{=}
{h_a}^\lambda {h^a}_{\mu, \nu}, \ee where $h_{a \mu, \
\nu}=\partial_\nu h_{a \mu}$.  The metric
tensor $g_{\mu \nu}$
 is defined by
 \be g_{\mu \nu} \stackrel{\rm def.}{=}  \eta_{a b} {h^a}_\mu {h^b}_\nu, \ee
with $\eta_{a b}=(-1,+1,+1,+1)$ is the  Minkowski
spacetime. The torsion  and the contorsion are defined as
\ba {T^\alpha}_{\mu \nu}  \A \stackrel {\rm def.}{=} \A
{\Gamma^\alpha}_{\nu \mu}-{\Gamma^\alpha}_{\mu \nu} ={h_a}^\alpha
\left(\partial_\mu{h^a}_\nu-\partial_\nu{h^a}_\mu\right),\quad {K^{\mu \nu}}_\alpha  \stackrel {\rm def.}{=}
-\frac{1}{2}\left({T^{\mu \nu}}_\alpha-{T^{\nu
\mu}}_\alpha-{T_\alpha}^{\mu \nu}\right)={\Gamma^\lambda}_{\mu
\nu}-\left \{_{\mu \nu}^\lambda \right\}. \ea  The tensor ${S_\alpha}^{\mu \nu}$ and the scalar tensor, $ T $, are
defined as \be {S_\alpha}^{\mu \nu} \stackrel {\rm def.}{=}
\frac{1}{2}\left({K^{\mu \nu}}_\alpha+\delta^\mu_\alpha{T^{\beta
\nu}}_\beta-\delta^\nu_\alpha{T^{\beta \mu}}_\beta\right),\qquad  T \stackrel {\rm def.}{=}
{T^\alpha}_{\mu \nu} {S_\alpha}^{\mu \nu}. \ee Similar to the
$f(R)$ theory, one can define the action of $f(T )$ theory as \be
{\cal L}({h^a}_\mu, \Phi_A)=\int d^4x
h\left[\frac{1}{16\pi}f(T)+{\cal L}_{Matter}(\Phi_A)\right],\ee
where $h=\sqrt{-g}$ and   $\Phi_A$ are the matter fields.
Assuming the action (5) as a functional of the fields
${h^a}_\mu$, $\Phi_A$. The vanishing of the variation  with respect to the field ${h^a}_\mu$ gives the
following equation of motion$^{ \rm {[14]}}$  \be {S_\mu}^{\rho \nu}
T_{,\rho} \
f(T)_{TT}+\left[h^{-1}{h^a}_\mu\partial_\rho\left(h{h_a}^\alpha
{S_\alpha}^{\rho \nu}\right)-{T^\alpha}_{\lambda
\mu}{S_\alpha}^{\nu
\lambda}\right]f(T)_T-\frac{1}{4}\delta^\nu_\mu f(T)=4\pi {\cal
T}^\nu_\mu,\ee where $T_{,\rho}=\frac{\partial T}{\partial
x^\rho}$, $f(T)_T=\frac{\partial f(T)}{\partial T}$,
$f(T)_{TT}=\frac{\partial^2 f(T)}{\partial T^2}$ and ${\cal
T}^\nu_\mu$ is the energy momentum tensor. In this study we will consider the
matter content to have anisotropic form, i.e., given by \be {\cal T}_{\mu
\nu}=diag(\rho,-p_r,-p_t,-p_t),\ee where, $\rho$, $p_r$ and
$p_t$ are the energy density, the radial and tangential pressures respectively. In
the next section we are going to apply the field Eq. (6) to a
spherically symmetric spacetime and try to find
  new solutions.

Assuming that the {\it non trivial} manifold possesses stationary and spherical
symmetry has the form$^{ \rm {[21]}}$  \be \left( {h^\alpha}_i \right)=
\left( \matrix{ e^{\frac{A(r)}{2}} & 0 & 0 & 0\vspace{3mm} \cr 0 &
e^{\frac{B(r)}{2}}\sin\theta\cos\phi &r \cos\theta\cos\phi & -r
\sin\theta\sin\phi \vspace{3mm} \cr 0
&e^{\frac{B(r)}{2}}\sin\theta\sin\phi &r \cos\theta\sin\phi & r
\sin\theta\cos\phi \vspace{3mm}  \cr 0
 & e^{\frac{B(r)}{2}}\cos\theta &-r \sin\theta & 0 \cr } \right),
 \ee where $A(r)$ and $B(r)$ are two unknown functions of $r$. The metric associated with (8) takes the form
 \[ds^2=-e^{A(r)}dt^2+e^{B(r)}dr^2+r^2(d\theta^2+\sin^2\theta d\phi^2).\] It is important to note that for the same metric and
the same coordinate basis, different frames result in different forms of equations of motion.$^{ \rm {[22]}}$
 Using (8), one can obtain $h = det ({h^a}_\mu) = e^{\frac{(A+B)}{2}}r^2\sin \theta$. With the use of Eqs. (3) and (4), one can obtain the
torsion scalar and its derivatives in terms of $r$ in the form \ba
T(r) \A=\A
-\frac{2\left(1-2e^{\frac{-B}{2}}-rA'e^{\frac{-B}{2}}[1-e^{\frac{-B}{2}}]+e^{-B}
\right)}{r^2}, \qquad where \qquad A'=\frac{\partial
A(r)}{\partial r}, \nonu
T'(r)\A=\A\frac{\partial T}{\partial r}=\frac{e^{-\frac{B}{2}}
\left(B'\left[r^2A'(2e^{-\frac{B}{2}}-1)+2r(e^{-\frac{B}{2}}-1)\right]+
2r\left[rA''-A'\right]\left[1-e^{-\frac{B}{2}}
\right]-4\left[2-e^{-\frac{B}{2}}-e^{\frac{B}{2}}\right]\right)}{r^3}.\nonu
\A \A \ea The field equations (6) for an anisotropic fluid have the
form \ba 4\pi \rho \A=\A
-\frac{e^{-\frac{3B}{2}}f_{TT}}{r^4}\Biggl(4\{[rA'-r^2A'']+rB'+3\}+3r^2A'B'+4e^{{B}}+e^{\frac{B}{2}}\left(r^2[2A''-A'B']-2r[B'+A']-12\right)\nonu
\A \A
+e^{-\frac{B}{2}}\left\{2r^2A''-2rA'-2rB'-2r^2A'B'-4\right\}\Biggr)-\frac{e^{-\frac{B}{2}}f_{T}}{2r^2}\left(2-rA'+
e^{-\frac{B}{2}}\left[rA'+rB'-2\right]\right) +\frac{f}{4},\nonu
 4\pi p_r \A=\A\frac{f_{T}}{2r^2}\left(e^{-\frac{B}{2}}[6+3rA'-4e^{\frac{-B}{2}}]-2e^{-\frac{B}{2}}\left[rA'+1\right]\right)
+\frac{f}{4},\nonu
 4\pi p_t \A=\A
-\frac{e^{-\frac{3B}{2}}f_{TT}}{4r^4}\Biggl(\{[8r^2A'B'(1+\frac{rA'}{8})+16rA'+8rB'-2r^3A'A''-8r^2A''+2r^2A'^2+24]\}+e^{\frac{B}{2}}\Biggl(4r[rA''\nonu
\A \A
-B'-2A']-2r^2B'A'-24\Biggr)+8e^{{B}}
+e^{-\frac{B}{2}}\Biggl\{4r^2A''-2r^2A'^2-8rA'-4rB'-6r^2A'B'-2r^3B'A'^2\nonu
\A \A
+2r^3A'A''-8\Biggr\}\Biggr)+\frac{f_{T}e^{-\frac{B}{2}}}{8r^2}\left(8-4e^{\frac{B}{2}}+e^{-\frac{B}{2}}\left[2rA'+2r^2A''+r^2A'^2-r^2A'B'-2rB'-4\right]\right)
+\frac{f}{4}.\ea  It is of
interest to note that tetrad (8) is used in (\cite{BMT}, Eq. (4$\cdot$2)).  Here we will find other solutions to Eq. (10)  without any assumption on the unknown function $B(r)$. In the next section we  go to find several solutions to Eq. (10) assuming some conditions on $f(T)$.\\
\centerline{\underline {First assumption: $T=constnt=T_0$}}
 From Eq. (9), it can be shown that $A(r)$ which satisfies $T=T_0$ and $T'=0$ has the form \be
A(r)=\frac{1}{2}\int \frac{\left[2-4e^{\frac{ -B(r)}{2}}+2e^{
-B(r)}+r^2T_0\right]e^{\frac{B(r)}{2}}}{r\left(1-e^{\frac{-
B(r)}{2}}\right)}dr+c_1,\ee where $c_1$ is a constant of
integration.   The weak and null energy conditions have the form \be \rho\geq 0, \qquad \rho+p_r\geq 0,
\qquad \rho+p_t \geq 0, \qquad \rho+p_r \geq 0, \qquad \rho+p_t \geq 0.\ee
The asymptotic form of energy density,  radial and  transversal
 pressures when $T=T_0$ are shown in figure 1.
\begin{figure}
\begin{center}
{\includegraphics[width=5cm]{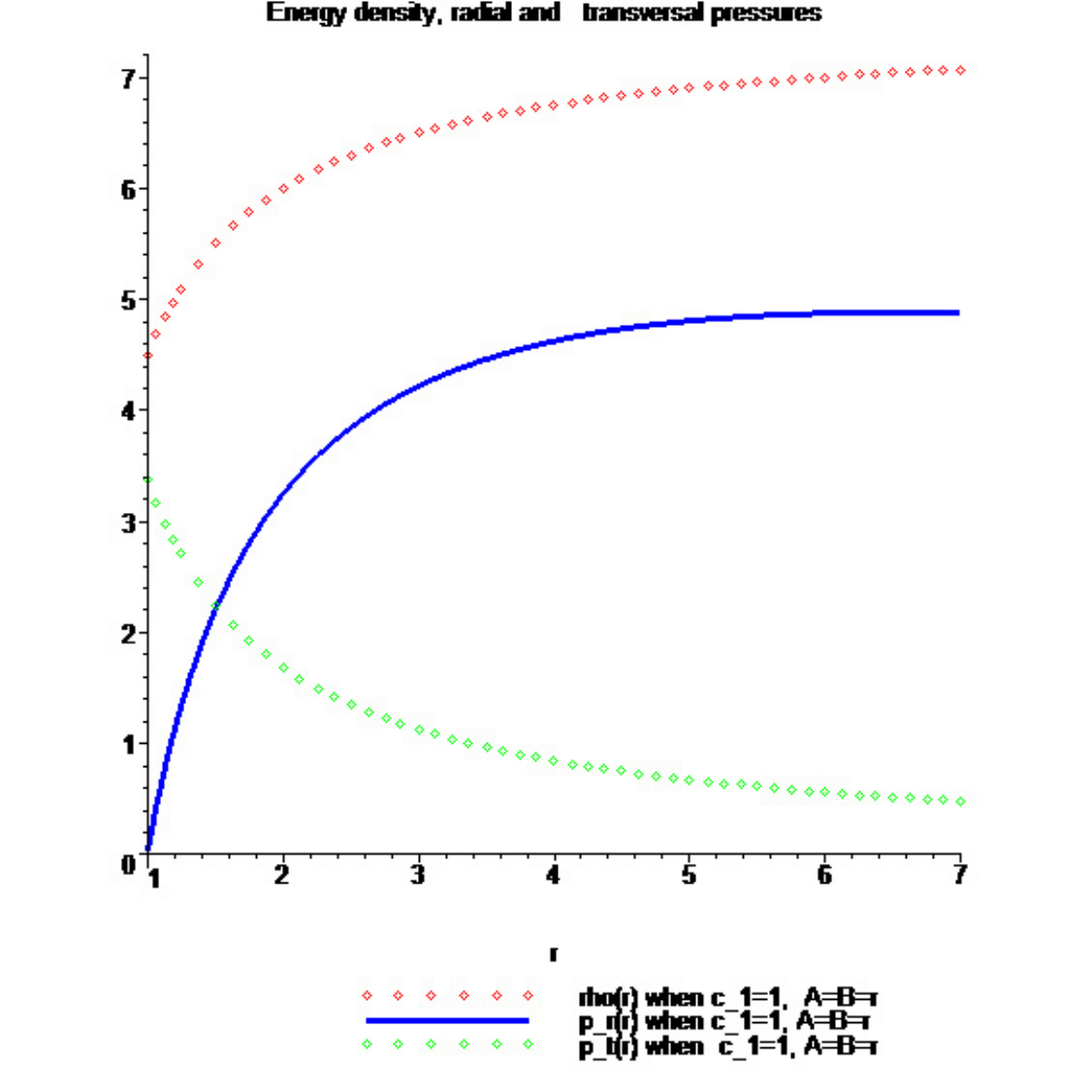}}
 \caption{\small{The energy density, radial and transversal pressures   when  $B(r)=r$,  $T_0=-4$,   $f(T_0)=9$ and $f_{T}(T_0)=3$.}}
\end{center}
\end{figure}
 From figure 1,
one can  show that the relevant physics  met to the standard energy
conditions, given by (12). We put
restrictions on the values of $T_0$, $f(T_0)$, $f_{T}(T_0)$ and
$B(r)$ such that Eq. (12) is satisfied.

In the case of vanishing  radial pressure the second of equations (10) reads \be f(T)=\frac{2f_T
e^{-\frac{B}{2}}}{r^2}\left(2+rA'-2e^{-\frac{B}{2}}\left[rA'+1\right]\right).\ee
We  go to study various cases of Eq. (13):\\
\centerline{\underline{ First form of $f(T)$}}
 In this case let us assume $f(T)$ to has the form$^{ \rm {[24]}}$ :
 \be
f(T)=a_0+a_1T+a_nT^n,\ee where $a_0$, $a_1$ and $a_n$ are
constants. From Eqs. (13) and (14) one can get for {\it the linear case
of $f(T)$} \be A(r)=\frac{1}{2}\int \frac{\left[e^{
B(r)}(2a_1-a_0r^2)-2a_1\right]}{ra_1}dr+c_2,\ee with $c_2$ being
another constant of integration.  The
asymptote  behavior of energy density  and the transversal
pressure, can be obtained from Eq. (10)  using Eq. (15), are shown
in  figure 2.
\begin{figure}
\begin{center}
{\includegraphics[width=5cm]{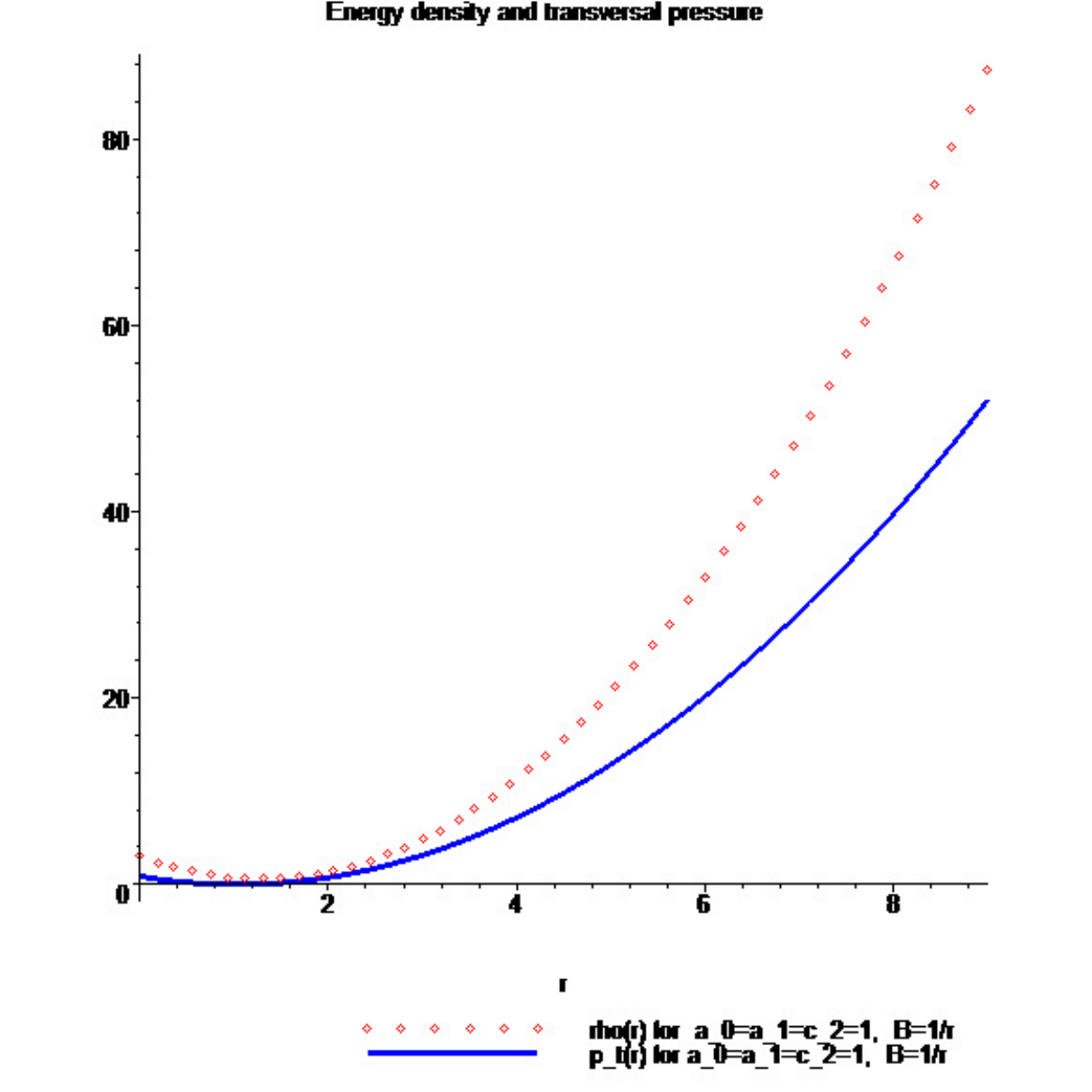}}
 \caption{The energy density and the tangential pressure   when   $a_0=a_1=c_2=1$, $B(r)\sim \frac{1}{r}$.}
\end{center}
\end{figure}
From figure 2, one can  show that the standard energy conditions
are satisfied which is  physically acceptable. This model is based on the asymptote of $B(r)\sim \frac{1}{r}$. Other
asymptote of $B(r)$ violate not only the standard energy
conditions but also causality.

 By the same way one can obtain a
solution for the {\it non-linear case}, i.e., when  $n=2$, in the form \ba A(r)\A=\A \frac{1}{4}\int
\frac{1}{r(3+e^{{B(r)}}
[1-4e^{\frac{-B(r)}{2}}])}\Biggl(e^{2B(r)}\left[e^{B}(a_1r^2+24a_2e^{\frac{B}{2}})-12a_2(1+e^{{B(r)}})\right]\nonu
\A \A
-\Biggl\{16{a_2}^2e^{{-3B(r)}}+16a_2e^{\frac{-5B}{2}}[a_1r^2-4{a_2}]
-e^{{-2B}}[12a_0a_2r^4-40a_1a_2r^2+{a_1}^2r^4+96{a_2}^2]\nonu
\A \A
+e^{\frac{-3B}{2}}[32a_1a_2r^2-16a_0a_2r^4-64{a_2}^2]+e^{{-B}}[4a_2a_0r^4-8a_2a_1r^2]+16{a_2}^2e^{\frac{-B}{2}}\Biggr\}^{1/2}\Biggr)
dr+c_3, \ea
with $c_3$ being constant of integration.
\begin{figure}
\begin{center}
{\includegraphics[width=5cm]{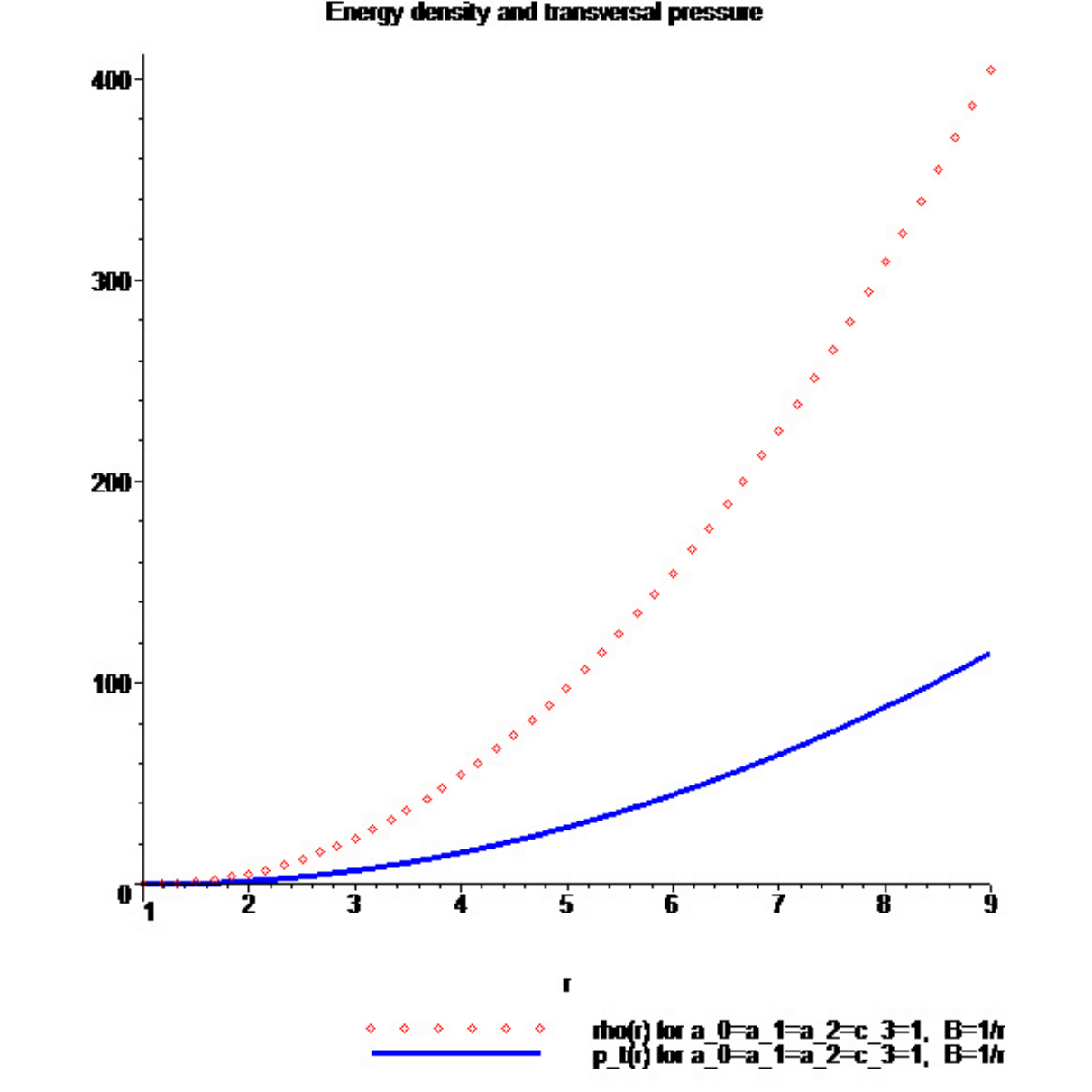}}
 \caption{The energy density and the tangential pressure   when  $a_0=a_1=a_2=c_3=1$, $B(r)\sim \frac{1}{r}$.}
\end{center}
\end{figure}
 From figure 3, it is clear that this model meets the
standard energy condition  when  $a_0=a_1=a_2=c_3=1$ and $B(r)\sim
\frac{1}{r}$. Other options are not permitted due to the reasons
discussed above.

 Using Eq. (13) in Eq. (14) we get \ba \A \A
\frac{1}{r^2}\Biggl\{[r^2\{a_0+a_1T+a_nT^n\}]-e^{\frac{-B(r)}{2}}[4a_1+2ra_1A'+2nra_nT^{n-1}A'+
4na_nT^{n-1}]\nonu
\A \A
+e^{-B(r)}[4na_nT^{n-1}+4ra_1A'+4a_1+4nra_nA'T^{n-1}]\Biggr\}=0.\ea
 According to the  model of $f(R)$, also used for the
$f(T)$ theory, as an alternative to the dark energy$^{ \rm {[14]}}$  by
taking the function $f(T)$  presented by Eq. (14) when $n=-1$, one
can obtain a relation between the unknown function $A(r)$ in terms
of the unknown function $B(r)$. The asymptote behavior of energy
density and the tangential pressure are plot in figure 4. Using
the same above procedure we get physically acceptable model as
shown in figure 4.\\
\begin{figure}
\begin{center}
{\includegraphics[width=6cm]{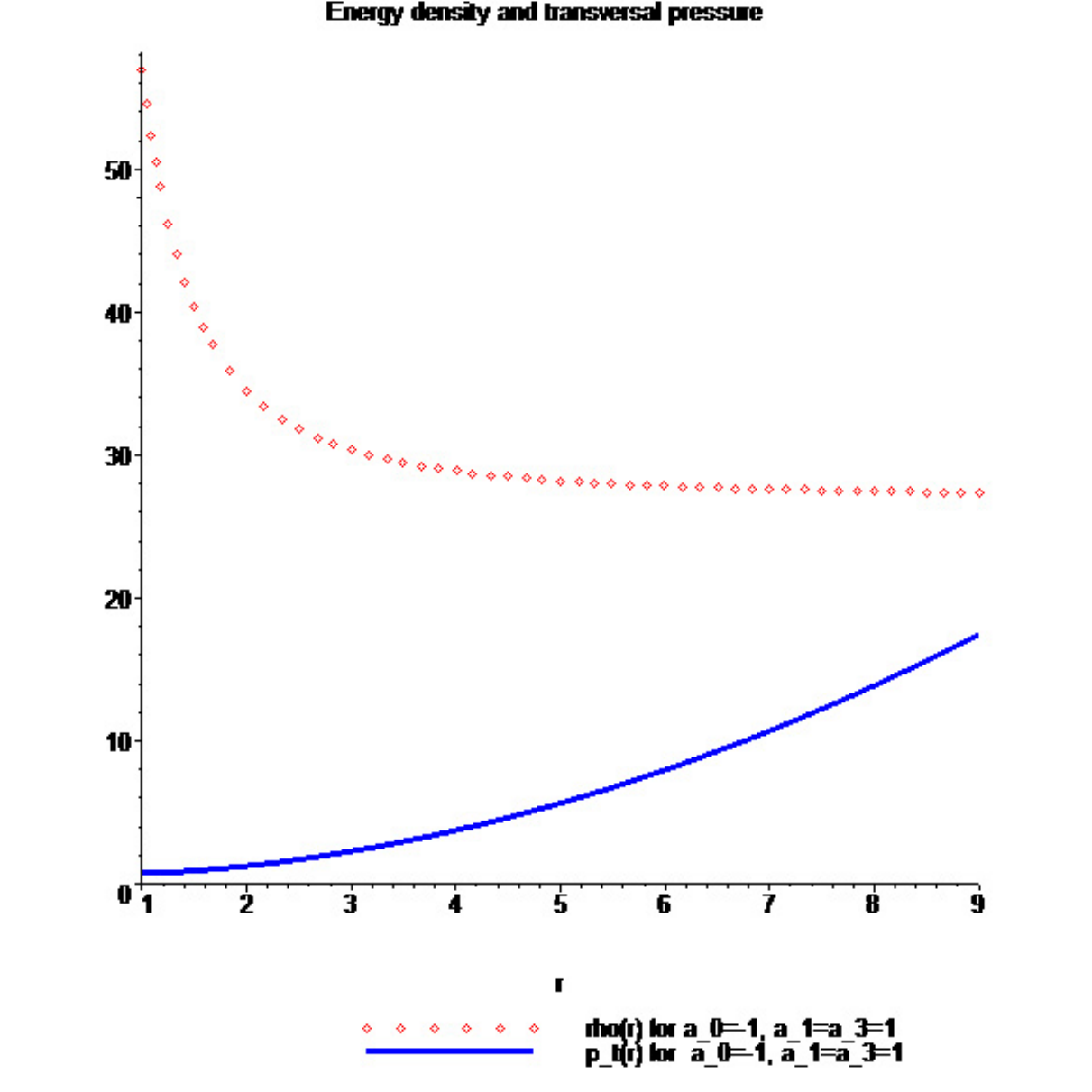}}
 \caption{The energy density and the tangential pressure   when   $a_0=a_1=a_2=1$ and $B(r)\sim \frac{1}{r}$.}
\end{center}
\end{figure}
\centerline{\underline{Main results and discussion}}
   All the  solutions  obtained in this work can be summarized as follow:\vspace{0.1cm}\\
  $\bullet$ When the scalar torsion is taken to be constant, i.e., $T=T_0$
   a quite general differential equation that governed the two
unknown functions  has been obtained.  By this relation
calculations of energy density, radial and transversal pressures
are provided. The asymptote behavior of these quantities are
drawn in figure 1 for specific asymptote of $B(r)$. From  figure
1, it  became clear that the asymptote of energy density, radial
and transversal pressures depend on the constants  as well as on
the asymptote behavior of $B(r)$.\vspace{0.1cm}\\
  $\bullet$   The condition of  vanishing  radial pressure
is derived. A quite general assumption on the from of $f(T)$ has been employed.  Different cases have been studied using this assumption:\vspace{.1cm}\\
i) When $n=0$, following the
procedure done in the case of constant scalar torsion,  a
differential equation that links the two unknown functions
$A(r)$ and  $B(r)$ has been derived.  The asymptote behavior of
the density and transversal  pressure
 are given in figure 2. From this figure
 one can show that both energy density and transversal pressure are positive.\vspace{.1cm}\\
ii) When  the condition $n=2$ is employed a solution is obtained
for a specific form of one of the two unknown functions. The
density and tangential pressure are given in figure 3. From this
figure one can conclude that the character of this model is physically acceptable since it satisfied the
standard energy conditions.\vspace{0.1cm}\\
iii) When  the condition $n=-1$, this case is studied for the $f(T)$
theory as an alternative to the dark energy$^{ \rm {[14]}}$  by taking
the function f(T) as given by Eq. (14), is employed
a solution is obtained for a specific form of one of the two
unknown functions. The density and tangential pressure are given
in figure 4. From this figure one can conclude that this model is
also  physically acceptable for the above discussed reasons.

\end{document}